\documentclass[pra,twocolumn,showpacs,floatfix,a4paper,superscriptaddress]{revtex4}
\usepackage{bm,color,graphicx,amsmath,txfonts}

\newcommand{\PP}{{\cal P}}

\begin{document}

\title{Measurement of complete and continuous Wigner functions for discrete atomic systems}

\author{Yali Tian}
\affiliation{State Key Laboratory of Quantum Optics and Quantum Optics Devices, Institute of Opto-Electronics, Shanxi University, Taiyuan 030006, China \\
and Collaborative Innovation Center of Extreme Optics, Shanxi University, Taiyuan 030006, China}
\author{Zhihui Wang}
\affiliation{State Key Laboratory of Quantum Optics and Quantum Optics Devices, Institute of Opto-Electronics, Shanxi University, Taiyuan 030006, China \\
and Collaborative Innovation Center of Extreme Optics, Shanxi University, Taiyuan 030006, China}
\author{Pengfei Zhang}
\affiliation{State Key Laboratory of Quantum Optics and Quantum Optics Devices, Institute of Opto-Electronics, Shanxi University, Taiyuan 030006, China \\
and Collaborative Innovation Center of Extreme Optics, Shanxi University, Taiyuan 030006, China}
\author{Gang Li}
\affiliation{State Key Laboratory of Quantum Optics and Quantum Optics Devices, Institute of Opto-Electronics, Shanxi University, Taiyuan 030006, China \\
and Collaborative Innovation Center of Extreme Optics, Shanxi University, Taiyuan 030006, China}
\author{Jie Li}\thanks{jieli6677@hotmail.com}
\affiliation{State Key Laboratory of Quantum Optics and Quantum Optics Devices, Institute of Opto-Electronics, Shanxi University, Taiyuan 030006, China \\
and Collaborative Innovation Center of Extreme Optics, Shanxi University, Taiyuan 030006, China}
\affiliation{Department of Physics, Zhejiang University, Hangzhou 310027, China}
\affiliation{Institute for Quantum Science and Engineering and Department of Biological and Agricultural Engineering, Texas A{\rm \&}M University, College Station, Texas 77845, USA}
\author{Tiancai Zhang}\thanks{tczhang@sxu.edu.cn}
\affiliation{State Key Laboratory of Quantum Optics and Quantum Optics Devices, Institute of Opto-Electronics, Shanxi University, Taiyuan 030006, China \\
and Collaborative Innovation Center of Extreme Optics, Shanxi University, Taiyuan 030006, China}

\begin{abstract}
We measure complete and continuous Wigner functions of a two-level cesium atom in both a nearly pure state and highly mixed states. We apply the method [T. Tilma {\it et al}., Phys. Rev. Lett. {\bf 117}, 180401 (2016)] of strictly constructing continuous Wigner functions for qubit or spin systems. We find that the Wigner function of all pure states of a qubit has negative regions and the negativity completely vanishes when the purity of an arbitrary mixed state is less than $\frac{2}{3}$. We experimentally demonstrate these findings using a single cesium atom confined in an optical dipole trap, which undergoes a nearly pure dephasing process. Our method can be applied straightforwardly to multi-atom systems for measuring the Wigner function of their collective spin state.
\end{abstract}


\date{\today}
\maketitle

\section{Introduction}

The Wigner function (WF)~\cite{Wigner}, originally introduced as a quantum analog of the classical phase-space distribution function, provides a powerful tool to represent quantum mechanics in phase space~\cite{Wigner2}. It is a quasiprobability distribution in that it acts like a probability distribution but can take negative values for some quantum states. The WF is originally designed for describing quantum systems with continuous degrees of freedom. It has been widely used, for example, in quantum optics to facilitate the visualization and tomographic reconstruction of quantum states~\cite{exp1,exp2,exp3,exp3b,exp4,exp4b,exp5,exp6,exp7,exp8}.

While it has been successfully applied in continuous variable (CV) systems, the generalizations of the WF to quantum systems with a finite-dimensional Hilbert space have proved challenging. Many efforts have been made along this line, which in general can be divided into two approaches based on the dimension, finite~\cite{Wootters,Leonhardt,Vourdas,Saraceno,Wootters2} or infinite~\cite{Takahashi,Agarwal,Dowling,Wolf,Brif,Luis,Klimov,Braunstein,Tilma,Tilma2,Glaser,Tilma3}, of the phase space, on which the WF is defined. Correspondingly, we refer to these two kinds as discrete and continunous WF, respectively. It remains an open question which approach is better. However, we note that a continuous WF for finite-dimensional systems seems more consistent with the original WF defined for CV systems. Unlike the gradual progress works~\cite{Takahashi,Agarwal,Dowling,Wolf,Brif,Luis,Klimov,Braunstein,Tilma,Tilma2}, which have their own restrictions either in the representation space or in the accuracy of representing the state, quite recently an elegant method~\cite{Tilma3} has been proposed for constructing {\it complete and continuous} WFs for spin or qubit systems. The method follows the displaced parity operator approach to defining the WF for CV systems~\cite{Wodkiewicz}. The key is, therefore, to find appropriate analogous displacement and parity operators for spin systems. By means of the Bloch sphere representation of the state of a qubit, both the displacement and parity operators have been properly defined satisfying all the requirements of the Stratonovich-Weyl correspondence~\cite{Tilma3}, and hence a {\it complete and continuous} WF has been strictly constructed for any two-level systems.

Continuous WFs have been measured for a collective spin state of an atomic ensemble~\cite{PhilippNat,Vuletic}. In Ref.~\cite{PhilippNat}, the WF is reconstructed using the inverse Radon transform implemented by a filtered back-projection algorithm~\cite{Natterer}. The method employed there does not guarantee positivity of the reconstructed density matrix in the presence of experimental noise~\cite{PhilippNJP}, which may become a crucial problem for quantitative studies. While Ref.~\cite{Vuletic} adopts the method of Ref.~\cite{Dowling}, with which a Wigner-like function is defined providing intuitively meaningful pictures, but it only works for systems of definite angular momentum (e.g., the totally symmetric subspace for an atomic ensemble and hence its phase space representation is not complete~\cite{Tilma2}), whereas Ref.~\cite{Tilma3} can handle arbitrary spin systems. Just recently, complete and continuous WFs have been measured for the first time for discrete systems of two Bell states and five-qubit Greenberger-Horne-Zeilinger state~\cite{TilmaIBM} based on IBM superconducting-qubit quantum processor~\cite{IBM}. Though convenient, using such a processor, the measurement of the WF suffers from various imperfections, such as indirect implementation of rotations and detection due to the limited operations that IBM has made available to the user, and considerable noises in the system resulting in imperfect operations and state preparation.

Adopting the WF defined in Ref.~\cite{Tilma3}, in this paper we measure complete and continuous WFs of a well controlled truly single two-level cesium atom. Unlike experiments involving a large number of atoms for quantum metrology~\cite{PhilippNat,Vuletic}, in which single-atom resolution is unavailable in both control and measurement, in our experiment a single cesium atom is controlled deterministically in a micro-sized dipole trap and undergoes a nearly pure dephasing process. We find that for an {\it arbitrary pure} state of a qubit its WF has always negative regions and the negativity vanishes if the purity of an {\it arbitrary mixed} state is less than $\frac{2}{3}$. We experimentally demonstrate these findings using our system of trapped single atoms. To our knowledge, this is the first time that {\it complete and continuous} WFs have been measured for discrete atomic systems and that the evolution of the corresponding WF in a dephasing environment has been demonstrated.

\section{Theory}

Any state of a two-level quantum mechanical system can be represented by a point on/in the Bloch sphere. The surface of the Bloch sphere represents all the pure states, whereas the interior corresponds to all the mixed states. Any Hermitian $2\times 2$ matrix $\rho$ with ${\rm tr} \rho\,{=}\,1$ can be expressed as $\rho=\frac{1}{2}(\mathbb{I} + \vec{r} \cdot \vec{\sigma} )$~\cite{ChuangBook}, where $\mathbb{I}$ is the identity matrix, $\vec{r}=r \vec{e}$ is the Bloch vector with magnitude $r$, $0\le r \le 1$, and unit vector $\vec{e}=(\sin\theta \cos\phi, \sin\theta \sin\phi, \cos\theta)$, which specifies a point on the surface of the Bloch sphere. $\theta$ and $\phi$ are the polar and azimuthal angle, respectively, $\theta\in[0,\pi]$ and $\phi\in[0,2\pi)$. $\vec{\sigma}$ is the 3-element `vector' of Pauli matrices $\vec{\sigma}=(\sigma_x, \sigma_y, \sigma_z)$. Thus, $\rho$ can be rewritten as
\begin{equation}
\rho(\theta,\phi,r)=\frac{1}{2}
\begin{pmatrix}
1+r \cos\theta & e^{-i \phi} r \sin\theta  \\
e^{i \phi} r \sin\theta  & 1-r \cos\theta
\end{pmatrix}.    
\label{eq2}
\end{equation}
Eq.~\eqref{eq2} denotes that any density matrix $\rho$ of a qubit can be characterized by the three parameters ($\theta,\, \phi$, $r$). The purity of the state is defined by $\PP \equiv {\rm tr}\rho^2=\frac{1}{2}(1+r^2)$. For pure states with $r=1$ purity $\PP=1$, while for mixed states with $0\le r<1$ purity $\frac{1}{2}\le\PP<1$. It is evident that the decreasing of $r$ from 1 to 0 corresponds to a decoherence process with the off-diagonal entries of $\rho$ decaying to zero.

We wish to simulate the decoherence process as $r$ decreases using our existing two-level atom system with the aim of observing the evolution of the corresponding WF defined in Ref.~\cite{Tilma3}. We notice that in general as $r$ decreases {\it all} the entries of $\rho$ vary corresponding to a complicated process that contains both dissipative and dephasing dynamics. However, for the special case of $\theta=\frac{\pi}{2}$, as $r$ reduces the diagonal entries of $\rho$ are left unchanged, i.e., $\rho_{11}=\rho_{22}=\frac{1}{2}$, and only the off-diagonal entries decay, corresponding to a pure dephasing process. This process can be accurately simulated using our system of single cesium atoms confined in an optical dipole trap. We shall explain this in more detail in the next section.

The continuous WF for such a two-level system is defined as~\cite{Tilma3}
\begin{equation}
W_{\rho}(\xi,\chi)={\rm tr}[\rho \hat{\Delta}(\xi,\chi)],
\label{eq3}
\end{equation}
with the operator $\hat{\Delta}(\xi,\chi)$ taking the form of
\begin{equation}
\hat{\Delta}(\xi,\chi)=\frac{1}{2}\Big[ \mathbb{\hat I} - \! \sqrt{3}\, \big( \hat{R} \, \hat{\sigma}_z \hat{R}^\dag \big) \Big],
\label{eq4}
\end{equation}
where $\mathbb{\hat I}$ is the identity operator, $\hat{\sigma}_z$ can be treated as the parity operator for a qubit, and $\hat{R}=e^{-i \frac{\xi}{2} \hat{\sigma}_z} e^{-i \frac{\chi}{2} \hat{\sigma}_x} e^{-i \frac{\Xi}{2} \hat{\sigma}_z}$ is the rotation operator that ``displaces" a qubit state along the surface of the Bloch sphere. $\xi$, $\chi$, and $\Xi$ are the Euler angles and it is known that any target orientation can be realized by composing three elemental rotations, i.e., rotations about the axes of the Bloch sphere. Note that $W_{\rho}(\xi,\chi)$ is a function of only two Euler angles ($\xi$, $\chi$) because $\Xi$ makes no contribution as $e^{-i \frac{\Xi}{2} \hat{\sigma}_z}$ commutes with $\hat{\sigma}_z$.

\begin{figure}[t]
\hskip-0.1cm\includegraphics[width=0.96\linewidth]{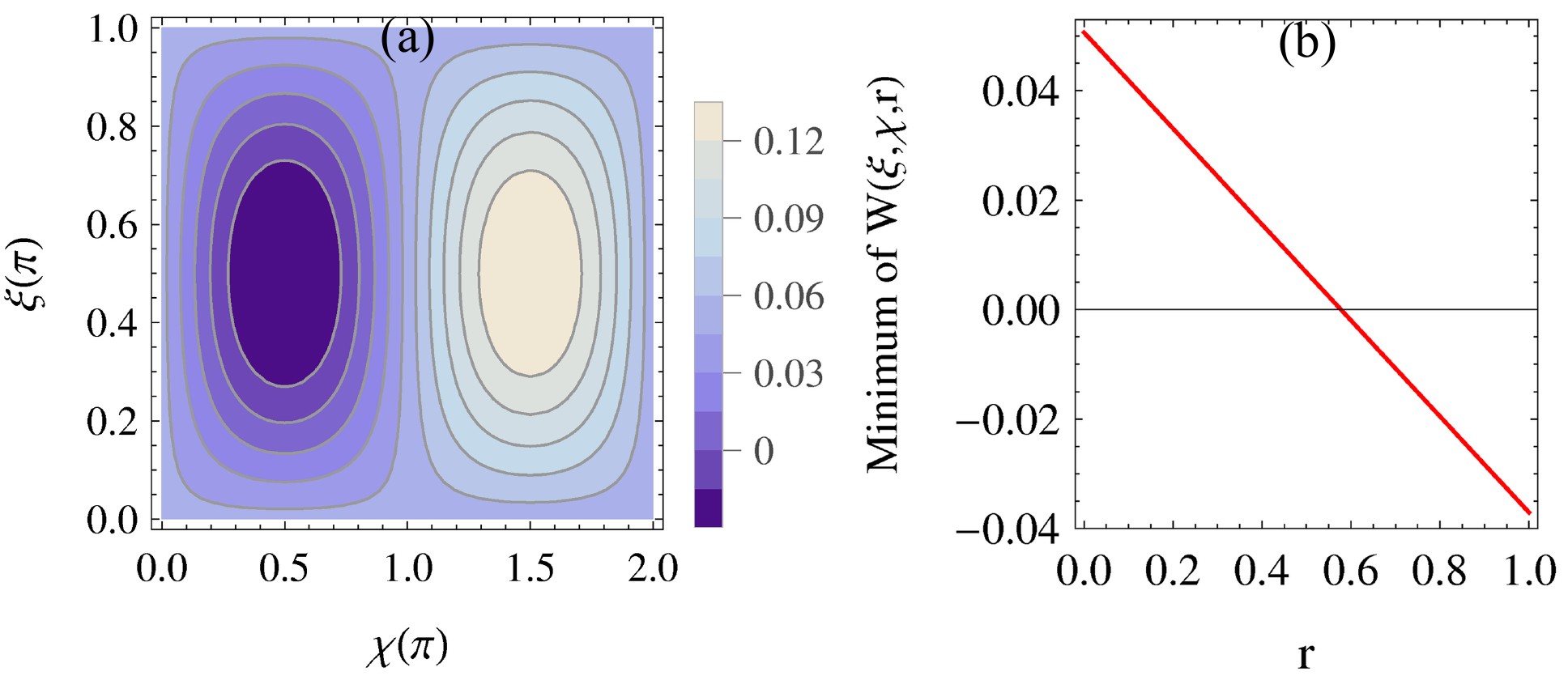}
\caption{(a) Theoretical WF $W(\xi,\chi)$ of the qubit state $(|0\rangle+|1\rangle)/\sqrt{2}$, i.e., of the state $\rho(\frac{\pi}{2}, 0, 1)$. (b) Minimum value of the WF $W(\xi,\chi; \theta, \phi, r)$ of a qubit (with arbitrary values of $\theta$ and $\phi$) versus $r$.}
\label{fig1}
\end{figure}

Inserting Eq.~\eqref{eq2} into Eq.~\eqref{eq3}, the WF for a generic qubit state $\rho(\theta,\phi,r)$ is therefore obtained  
\begin{equation}
\begin{split}
&W(\xi,\chi;\theta,\phi,r)=   \\
&\,\,\,\,\,\,\,\,\,\frac{1}{2\pi^2} \bigg\{1{-}\sqrt{3} r \, \Big[ \! \cos\theta \cos \chi + \sin(\xi-\phi) \sin\theta \sin \chi  \Big]  \bigg\},
\label{eq5}
\end{split}
\end{equation} 
where $\frac{1}{\pi^2}$ is introduced to make the WF normalized over the phase space $\xi \in [0,\pi]$ and $\chi \in [0, 2\pi)$. We note that both $\xi$ and $\chi$ have a period of $2\pi$, however, a space of half a period of $\xi$ and a period of $\chi$ is enough to determine a WF that contains complete information of the state. It is straightforward to check that $W(\xi,\chi, r)$ is in all regions positive when $r<\frac{1}{\sqrt{3}}$, or when purity $\PP<\frac{2}{3}$, since the sum of the two trigonometric terms is bounded by $\pm 1$. This is a general result for a qubit state of arbitrary values of $\theta$ and $\phi$. Besides, for all pure states ($r{=}1$) the WFs always have negative regions and, interestingly, they possess the same minimum value $W_{\rm min}\,{=}\,\frac{1}{2\pi^2}(1{-}\sqrt{3})\,{\approx}\, {-}0.037$. In CV systems, the negativity of the WF is typically considered as a nonclassical signature of the state~\cite{VogelBook,UlfBook}. However, in discrete systems things are more complicated because the negativity shows subtle complexities~\cite{TilmaIBM}. While for more general mixed states, the minimum is only related to $r$ regardless of $\theta$ and $\phi$, i.e., $W_{\rm min}\,{=}\,\frac{1}{2\pi^2}(1{-}\sqrt{3}r)$. There exists a critical value of $r\,{=}\,\frac{1}{\sqrt{3}}\, {\simeq} \,0.577$ (or of $\PP=\frac{2}{3}$), below which the negative regions of the WF completely vanish. This is clearly shown in Fig.~\ref{fig1}.

\section{Experimental setup and procedures}

To make the WF Eq.~\eqref{eq3} more closely linked to the actual operations in an experiment, we rewrite it as
\begin{equation}
W_{\rho}(\xi,\chi)=\frac{1}{2\pi^2} \Big[ 1- \sqrt{3}\, {\rm tr}(\rho' \hat{\sigma}_z ) \Big],
\label{eq6}
\end{equation}
where $\rho'=\hat{R}_x(-\chi) \hat{R}_z(-\xi)  \rho \hat{R}_z^{\dag}(-\xi)  \hat{R}_x^{\dag}(-\chi)$, and $\hat{R}_z(\xi)=e^{-i \frac{\xi}{2} \hat{\sigma}_z}$ and $\hat{R}_x(\chi)=e^{-i \frac{\chi}{2} \hat{\sigma}_x}$ correspond to the rotation about the $z$ and $x$-axis of the Bloch sphere, respectively. Eq.~\eqref{eq6} denotes that the WF of $\rho$ is connected to the expectation value of $\hat{\sigma}_z$ over the state $\rho'$ that is achieved by performing two sequential rotation operations on $\rho$. To be more intuitive, we express Eq.~\eqref{eq6} in an equivalent form  
\begin{equation}
W_{\rho}(\xi,\chi)=\frac{1}{2\pi^2} \Big[ 1- \sqrt{3} (P_0 - P_1) \Big],
\label{eq7}
\end{equation}
where $P_0=\langle 0|\rho'|0 \rangle$ and $P_1=\langle 1|\rho'|1 \rangle$ are, respectively, the population probability of the two eigenstates $|0\rangle$ and $|1\rangle$. In our system, these two states are embodied by the ``clock states" of a cesium atom, i.e., $|0\rangle \equiv |6 S_{1/2}, F{=}3, m_F{=}0\rangle$ and $|1\rangle \equiv |6 S_{1/2}, F{=}4, m_F{=}0\rangle$~\cite{BohrPRA}.

\begin{figure}[t]
\hskip-0.25cm\includegraphics[width=0.95\linewidth]{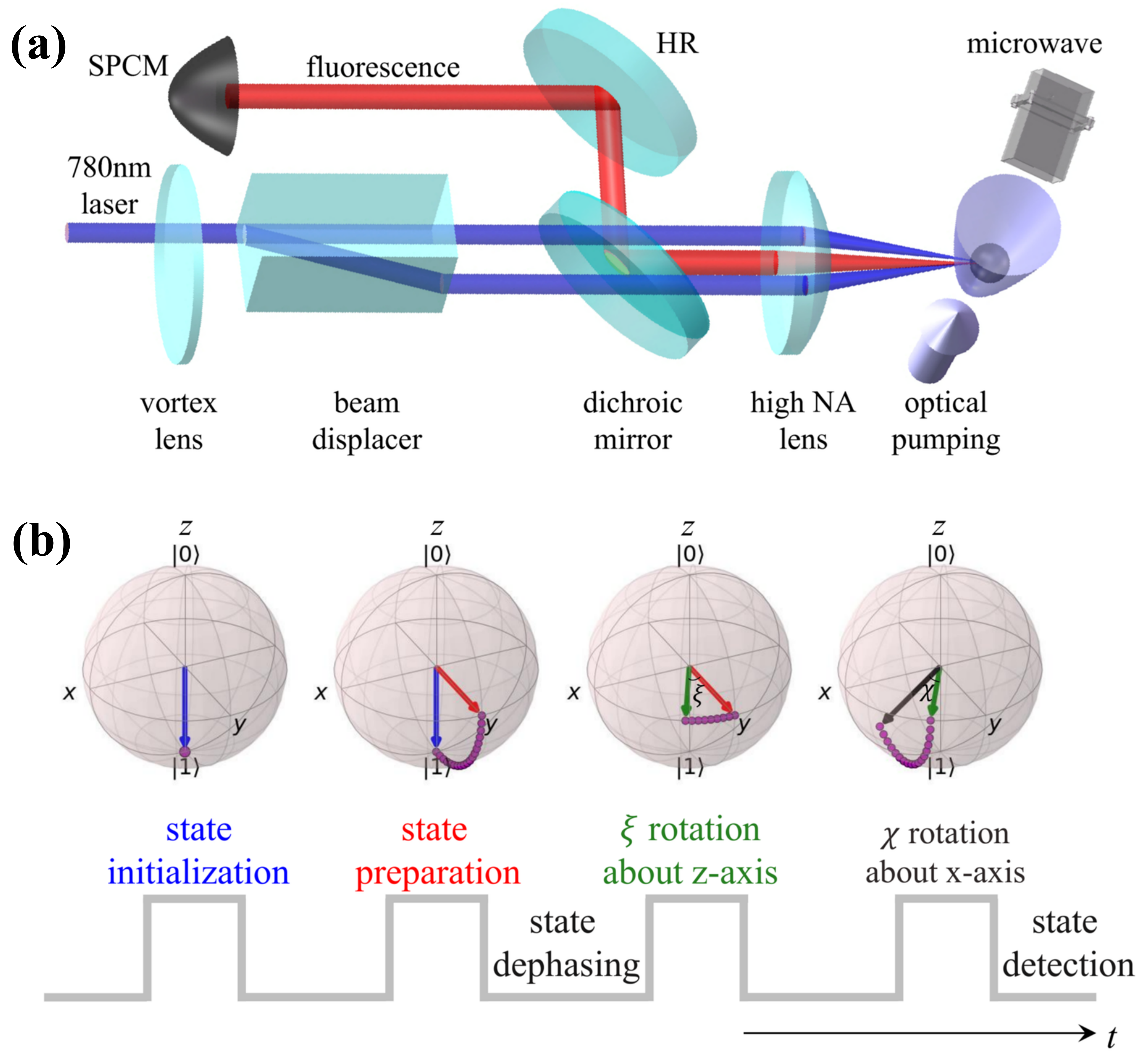}
\caption{(a) Experimental setup and (b) operation sequence for the measurement of the WF of a qubit.}
\label{fig2}
\end{figure}

The experimental setup is depicted in Fig.~\ref{fig2}(a). Single cesium atoms are repeatedly captured with a blue-detuned ``bottle" beam trap~\cite{ligang,BohrPRA}, which is superposed with a precooled atomic ensemble prepared by a conventional magneto-optical trap (MOT)~\cite{09PRA}. The ``bottle" trap is formed by shining two parallel ``donut" 780 nm laser beams with orthogonal polarizations through a group of high numerical aperture (NA) lens. By properly designing the size of the ``bottle" trap, no more than one atom at a time could be loaded from the MOT into the trap~\cite{BohrPRA}. The trapped atom is cooled to a temperature ${\sim} 10$ $\mu$K by polarization gradient cooling. The scattering photons by trapped single atoms are collected and eventually fed to a single photon counting module (SPCM). A microwave is nearly resonant with the 9.2 GHz hyperfine transition of the two ``clock states" and is applied to perform the corresponding operation on the qubit. The microwave generator is locked to a commercial Rb atomic clock to stabilize the frequency of the microwave.

The sequence of the operations is shown in Fig.~\ref{fig2}(b). A single trapped atom is initialized to state $|1\rangle$ by optical pumping. Then a microwave pulse is used to prepare the atom into a superposition state $|\psi\rangle=\cos \frac{\theta}{2} |0\rangle+e^{i\phi} \sin \frac{\theta}{2} |1\rangle$. In order to verify the state that has been prepared, one needs to do state tomography of the atomic density matrix. This process is of nonnegligible time (about 1 ms) and will make the superposition state evolve into a slightly mixed state with purity close to unity. It has been shown that in such an optical dipole trap the atom suffers from a pure dephasing mechanism~\cite{Kuhr,FrontPhys}. This fact has been verified by making state tomography at different decoherence time (see Appendix A). We have explained previously that the only situation corresponding to a pure dephasing process as $r$ reduces is that the initial state should be prepared with $\theta \simeq \frac{\pi}{2}$~\cite{note}.

\begin{figure}[b]
\hskip0cm\includegraphics[width=0.95\linewidth]{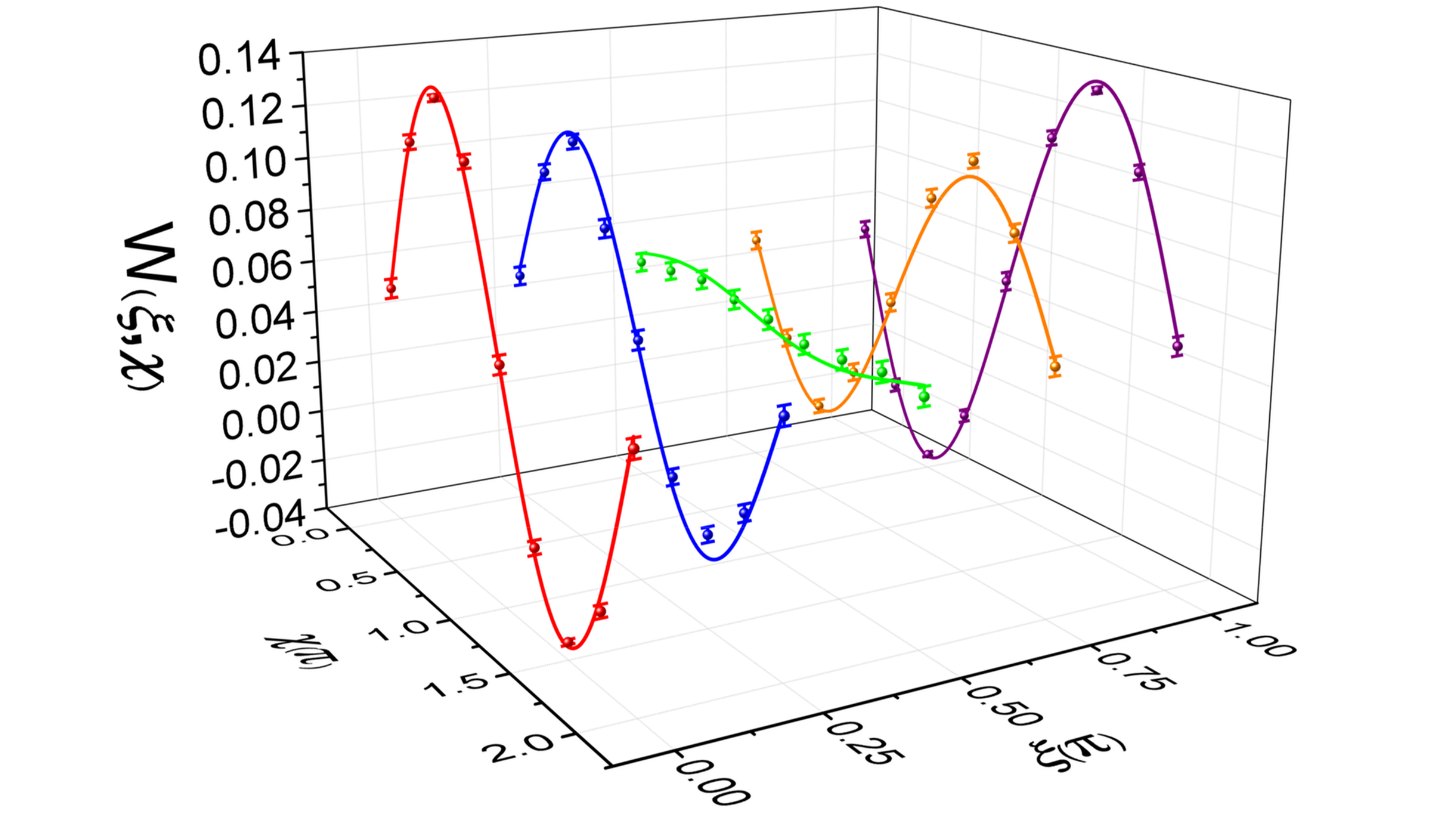}  \\
\caption{Experimental and theoretical WF of a qubit. Dots (with error bar): experimentally measured $W(\xi,\chi)$ of the state $\rho_{0}$ (see text). Curves: theoretically evaluated $W(\xi, \chi)$ [using Eq.~\eqref{eq5}] of $\rho(0.509\pi, 0.521\pi, 0.981)$, which has a unity fidelity with $\rho_{0}$. Curves from left to right correspond to $\xi=0$, $\frac{\pi}{4}$, $\frac{\pi}{2}$, $\frac{3\pi}{4}$, and  $\pi$, respectively.}
\label{fig3}
\end{figure}

\begin{figure*}[t]
\hskip-0.38cm\includegraphics[width=0.88\linewidth]{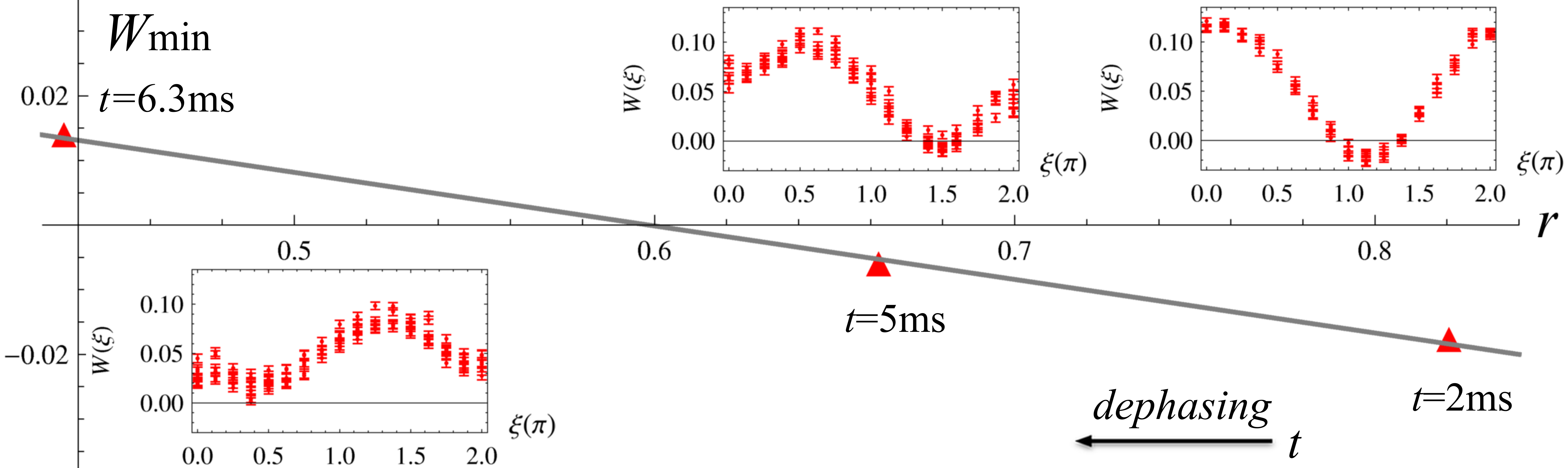}
\caption{Measured minimum value $W_{\rm min}$ versus $r$ in the dephasing channel. The insets are the experimental WF $W(\xi)$ at $\chi=\frac{\pi}{2}$ of three mixed states at time $t=2$ ms, 5 ms, and 6.3 ms, respectively, starting from the initial state $\rho_0$. The gray line is the fit line of $W_{\rm min}(r)$ based on the average values of $W_{\rm min}$ and $r$ at the three different time. }
\label{fig4}
\end{figure*}

After the stage of state preparation, the atom evolves through a dephasing channel for a time $t$, and then the stage of measurement of the WF starts. It is comprised of three sequential operations: two rotations and a detection (see Fig.~\ref{fig2}(b)). Specifically, a series of microwave pulses are used to implement rotations about the $z$ and $x$-axis of the Bloch sphere. Rotation of $\xi$ about $z$-axis can be controlled by the fact that $\xi=t\,\Delta$, with $\Delta$ the detuning of the microwave from the transition frequency of the two eigenstates. While rotation of $\chi$ about $x$-axis can be implemented by acting on a microwave pulse for a time $t=\frac{\chi}{\Omega_R}$, with $\Omega_R$ the Rabi flopping frequency associated with the two ``clock states". Finally, we measure the population probabilities $P_0$ and $P_1$ of the states $|0\rangle$ and $|1\rangle$, respectively. To this end, we adopt the method of Ref.~\cite{Kuhr2}, i.e., to push the atom in $|1\rangle$ out of the dipole trap by sending another laser beam, whereas the atom in $|0\rangle$ remains trapped. By checking if the atom still stays in the trap, one can discriminate in which state the atom is. After repeating the experiment many times, one then gets the population probabilities $P_0$ and $P_1$. Therefore, a value of the WF is achieved according to Eq.~\eqref{eq7} for specific rotations of $\xi$ and $\chi$. Repeating the experiments for different values of $\xi$ and $\chi$, a 3D WF $W_{\rho}(\xi,\chi)$ of the state $\rho$ at time $t$ could be measured for the whole phase space. Note that in practice the measured WF is the representation of the state at time $t+t_m$, with $t_m$ the measurement time which is less than 1 ms (specifically depending on the rotation angle) and much shorter than the atomic coherence time ${\sim} 17.2$ ms (see Appendix B).

\section{Results and discussion}

Figure~\ref{fig3} presents the experimental WF for the state $\rho_{0}$ prepared at the initial time. The entries of $\rho_{0}$ are measured via state tomography and each entry is obtained by the statistic of about 300 rounds of the measurement: $\rho_0^{11}{=}0.486 {\pm} 0.020$, $\rho_0^{22}{=}0.514 {\mp} 0.020$, $\rho_0^{12,21}{=}\,(-0.033 \pm 0.020) \mp (0.489 {\pm} 0.004) i$, corresponding to purity $\PP \,{\simeq}\, 0.981$ and $r \,{\simeq}\, 0.981$. The initial state $\rho_{0}$ is of $\theta \, {\simeq}\, (0.509 \pm 0.013)\pi$, which is very close to the desired state of $\theta\,{=}\,\frac{\pi}{2}$. The state $\rho_{0}$ (taking average values of its entries) has a unity fidelity with the state $\rho(0.509\pi, 0.521\pi, 0.981)$. In Fig.~\ref{fig3}, each dot with error bar is obtained by the statistic of about 300 times of the measurement and the curves are the theoretical WF of $\rho(0.509\pi, 0.521\pi, 0.981)$ for a series of values of $\xi$. It shows that the experimentally measured values are in good agreement with the theoretical curves. The small difference between the experimental and theoretical WFs is the result of many factors, such as the difference of measurement time of the WF and state tomography (based on which we obtain $\rho_{0}$ and plot the theoretical curves), and the non-unity contrast of Rabi flops (about 90\% as shown in the figure in Appendix B) which affects the fidelity of rotation operations and thus the accuracy of the measured WF.

As the state evolves in the dephasing channel, the state becomes more and more mixed (with a decreasing $r$) and the phase $\phi$ will have an increasing fluctuation, leading to an increasing uncertainty of the WF in $\xi$. In Fig.~\ref{fig4} (insets), we present experimental WFs of three mixed states at different evolution time. We have measured the WF for a period of $\xi$ at $\chi=\frac{\pi}{2}$ and then the minimum value will be of high possibility within the range $\xi \in [0, 2\pi)$. This is because the initial state $\rho_{0}$ of $\theta \simeq \frac{\pi}{2}$ guarantees the minimum value be at (or very close to) $\chi{=}\frac{\pi}{2}$. In each inset, the corresponding value of $r$ is achieved by the ensemble average of more than 10 times state tomography (each of which yields a value of $r$) at the same time:  $r=0.820^{+0.104}_{-0.137} $ at $t{=}2$ ms; $r=0.662^{+0.091}_{-0.153} $ at $t{=}5$ ms; and $r=0.436^{+0.099}_{-0.154} $ at $t{=}6.3$ ms. The fluctuation of $r$ at the same time is due to the fluctuation of the phase embodied by the considerable differences of the off-diagonal entries at different times of tomography. The insets of Fig.~\ref{fig4} show clearly that the width (reflecting fluctuation) of the Wigner ``stripe" increases with the evolution time as a result of an increasing fluctuation in the phase. The mismatch of the WF at $\xi =0$ and $2\pi$ is due to the nonnegligible time (less than 1 ms) of the $z$ rotation operation.

As shown previously, the minimum of the WF is connected to $r$ by $W_{\rm min}{=}\frac{1}{2\pi^2}(1{-}\sqrt{3}r)$. Despite a considerable fluctuation of $r$, it is still possible to verify the formula with {\it average} values of $r$ and $W_{\rm min}$ achieved by many times of measurements. In the insets of Fig.~\ref{fig4}, the averages of $r{=}0.820$, 0.662, 0.436 yield averages of $W_{\rm min}{=}-0.021$, $-0.007$, and $0.012$, respectively, by the formula. While the averages of more than 10 times measured $W_{\rm min}$ are $-0.018$, $-0.006$, and $0.014$, respectively, which are in good agreement with the values evaluated by the formula. The fit line of $W_{\rm min}(r)$ in Fig.~\ref{fig4} demonstrates the ``negative-to-positive" transition of the WF about $r \, {\simeq}\, 0.577$, or purity $\PP \, {\simeq} \,\frac{2}{3}$, almost perfectly verifying the theoretical expectations of Fig.~\ref{fig1}(b). We note that the measured $W_{\rm min}$ at $\chi=\frac{\pi}{2}$ is actually a bit higher than the ``real" $W_{\rm min}$ since the initial state $\rho_0$ is prepared {\it not exactly} at $\theta=\frac{\pi}{2}$. This makes the fit line move upwards a bit, leading to the intersection with $W_{\rm min}=0$ a bit larger than $r \simeq 0.577$.

\section{Conclusions}

We have measured complete and continuous WFs of a single two-level cesium atom in both a nearly pure state and highly mixed states following the method of Ref.~\cite{Tilma3}. We have shown how the WF evolves in a dephasing channel and demonstrated the ``negative-to-positive" transition when the purity of the state is about $\frac{2}{3}$. Our approach can in principle be applied to measure WFs of any two-level systems, either for a single qubit or for many qubits by implementing identical rotations on each qubit~\cite{TilmaIBM} still allowing obtaining a visible 3D WF at the price of losing partial information of the state. Furthermore, the demonstration of the WF evolving in a dephasing channel provides a more intuitive phase-space approach to studying fundamental processes in quantum discrete systems, such as the dynamics of decoherence.

\section*{ACKNOWLEDGMENT}

We would like to thank T. Tilma, H. Shen and T. Xia for fruitful discussions. This work has been supported by the National Key Research and Development Program of China (Grant No. 2017YFA0304502), the National Natural Science Foundation of China (Grants No. 11634008, No. 11674203, No. 11574187, and No. 61227902) and the Fund for Shanxi ``1331 Project" Key Subjects Construction.

\section*{Appendix A: State tomography for verifying the nearly pure dephasing process}

In our system, a single cesium atom is confined in an optical dipole trap, which undergoes a nearly pure dephasing process~\cite{Kuhr,FrontPhys}. In what follows, we further verify this fact by making state tomography at different time in this process. This is necessary since it provides a way for estimating the value of $r$ which is a key parameter in our model and it is also helpful to understand the physics of this process. The pure dephasing nature is characterized by the unchanged diagonal entries and the decaying off-diagonal ones of the density matrix as the state evolves. In Table I, we present density matrices measured at different time in the decoherence channel. We see that in this process the diagonal entries are almost unchanged, about 0.5, with consideration of measurement errors, while the off-diagonal ones may vary significantly and decay with the time. This is a clear signature of (nearly) pure dephasing in such a decoherence process.

\begin{table*}
\begin{center}
\caption{Density matrices of the qubit state measured in the dephasing channel. The purity and $r$ are calculated based on the density matrices taking the average value of their entries.}
\begin{tabular}{ p{1.6cm}  p{9.8cm}   p{1.4cm}  p{1.2cm} }    \hline 
Time (ms)   &  $\,\,\,\,\,\,\,\,\,\,\,\,\,\,\,\,\,\,\,\,\,\,\,\,\,\,\,\,\,\,\,\,\,\,\,\,\,\,\,\,\,\,\,\,\,\,\,\,\,\,\,\,\,\,\,\,\,\,\,\,\,\,\,\,\,\,\,\, {\rm Density \,\, matrix}$   &   Purity   &  $\,\,\,\, r$   \\ \hline
$\,\,\,\,\,\,0$    &   $\begin{pmatrix}
0.486 \pm 0.020   &  -0.033 \pm 0.020 -i (0.489 \pm 0.004)  \\
 -0.033 \pm 0.020 +i (0.489 \pm 0.004)   &  0.514 \mp 0.020  \\
\end{pmatrix}$      &   0.981   &  0.981    \\  \hline

$\,\,\,\,\,\,2$    &   $\,\,\, \begin{pmatrix}
0.503 \pm 0.020   &   0.207 \pm 0.018 -i (0.330 \pm 0.017)  \\
0.207 \pm 0.018 +i (0.330 \pm 0.017)   &  0.497 \mp 0.020  \\
\end{pmatrix}$       &   0.804   &  0.779    \\  \hline

$\,\,\,\,\,\,5$    &    $\,\,\, \begin{pmatrix}
0.507 \pm 0.021   &   0.321 \pm 0.017 -i (0.085 \pm 0.020)  \\
0.321 \pm 0.017 +i (0.085 \pm 0.020)   &  0.493 \mp 0.021  \\
\end{pmatrix}$       &   0.721   &  0.664    \\  \hline

$\,\,\,\,6.3$ &     $\begin{pmatrix}
0.507 \pm 0.019   &   -0.200 \pm 0.018 +i (0.135 \pm 0.019)  \\
-0.200 \pm 0.018 -i (0.135 \pm 0.019)   &  0.493 \mp 0.019  \\
\end{pmatrix}$       &   0.617   &  0.483    \\  \hline
\end{tabular}
\end{center}
\end{table*}

\begin{figure}[b]
\hskip-0.3cm\includegraphics[width=0.85\linewidth]{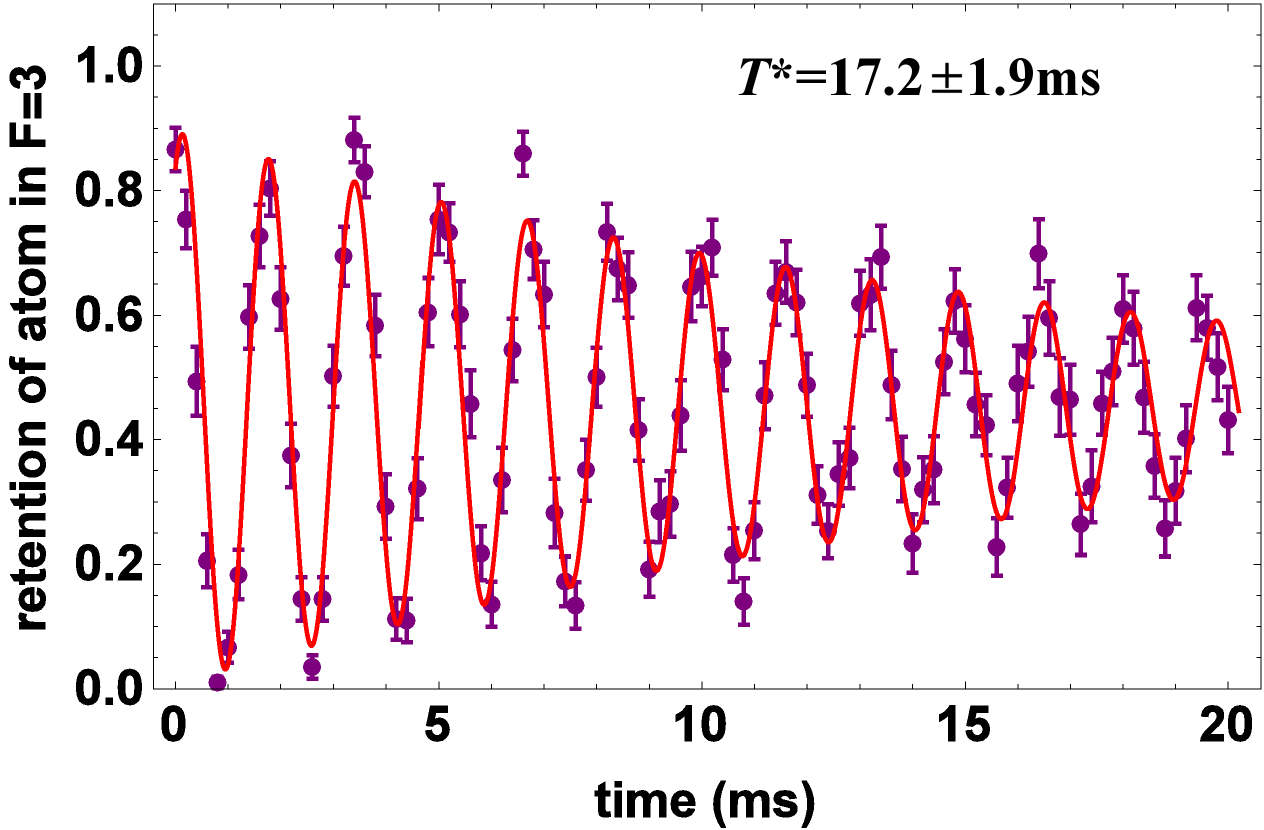}  \\
\caption{Ramsey fringes versus time delay between the two $\pi/2$ pulses. Dots (with error bar): experimental data and each dot is obtained by the statistic of about 100 times measurements. Curve: the fit curve of the Ramsey oscillation.}
\label{Ramsey}
\end{figure}

\section*{Appendix B: Estimation of the coherent time of the single qubit by Ramsey interference}

Here we briefly discuss the details of the approach to estimating the coherent time of the qubit in our experiment. The qubit is encoded in the ``clock states" of a cesium atom, i.e., $|0\rangle \equiv |6 S_{1/2}, F{=}3, m_F{=}0\rangle$ and $|1\rangle \equiv |6 S_{1/2}, F{=}4, m_F{=}0\rangle$. Firstly, the qubit is initialized to state $|1\rangle$ and then a resonant microwave pulse at frequency $9.2$ GHz is applied to drive the Rabi flopping. By using single atom Ramsey interferometry~\cite{BohrPRA,FrontPhys}, the coherent time $T^*$ can be precisely measured. A $\pi/2$ pulse is used to prepare the atom into the superposition state $(|0\rangle+|1\rangle)/\sqrt{2}$. After a time $t$, during which the state evolves freely in the far-off resonance trap~\cite{Ovchinnikov}, a second $\pi/2$ pulse is applied and then the state detection is performed. Fig.~\ref{Ramsey} shows the Ramsey interference signal of the atom versus the time interval $t$. The amplitude damping follows an exponential decay and the exponential fitting gives a $1/e$ decay time of $T^* \sim 17.2 \pm 1.9$ ms, that is the coherence time of the superposition state embodied in the atom. In our system, the temperature of the atom is about 10 $\mu$K measured using the method of release and recapture~\cite{Tuchendler}. The main factor of dephasing is due to the atom motion induced inhomogeneous dephasing~\cite{Kuhr}. This result indicates that the tomography measurement process of about 1 ms is much shorter than the coherent time of the state, which offers the possibility to precisely measure the qubit state at any evolution time.


\begin{thebibliography}{99}
\bibitem{note}
Note that no matter the atom is prepared in the state of any value of $\theta$, the atom suffers only pure dephasing, as shown in Refs.~\cite{Kuhr,FrontPhys}. However, only for the special case of $\theta=\frac{\pi}{2}$, we can model such a process with a simple reduction of $r$.
\end{thebibliography}
\end{document}